\documentclass[12pt]{article}

\begin{document}

\author{Dario Sassi Thober \and Center for Research and Technology CPTec - Unisal \and thober@fnal.gov thober@cptec.br \and Phone:55-19-7443107 Fax:55-19-7443036}
\title{The monopoles in the structure of the electron }
\date{}
\maketitle

\begin{abstract}
The classical electron is presented as made up of an electric charge and two
Dirac monopoles of opposite charge performing a magnetic dipole. It is
discussed that a valid variational principle for this system can be defined. The
Dirac quantization condition for charges can be derived in Quantum Electrodynamics as monopoles are properly defined in the structure. Further comments on the subject are addressed since the proper care of Dirac strings is still an open issue. 
\end{abstract}

\section{Introduction}

The monopole idea as derived by Dirac [1] brought a natural expalnation to
the charge quantization in Electrodynamics. The existence of a single
monopole in Nature could explain the discretization of all particles electric charges.
After the initial idea the monopole concept faced some difficulties to be
described through an action principle, [2,3,4]. As the action principle for
particles (where the monopole is considered as a source of magnetic field)
and fields could not be properly defined, the quantization of charge in
Electrodynamics as proposed by Dirac [1] turned out to be problematic. In
fact even in the Fiber Bundle formalism, [6], the problem persists because
the definition of a monopole in Quantum Electrodynamics suffers of two
theoretical problems: The lack of a general variational principle and the
proper care of the singular Dirac strings. The work of Wu and Yang, [6],
attempts to solve the second only.

The objective of the present work is to propose an electromagnetic structure
for the electron (or any other spin particle) as composed by an electric
charge and two Dirac monopoles (these last two being of opposite magnetic
charge). It is possible to show that an action principle can be assigned for
this system, in which both the particles and fields can be derived. As a
classical system, it can be shown also that, contrary to the primer
classical investigations, the size of the electron's structure must not to
be of the order of the electron's Compton wave lenght (in fact it may have a
size as small as desidered). Another interesting consequence is that, being
possible to describe a particle's spin by monopoles in its structure, the
quantization of charges proposed by Dirac holds in Quantum Electrodynamics.
\par In view of recent contributions on the Dirac string problem, [7], it is interesting to consider the present work in this context, which it is addressed at the conclusion section.

\section{The electron's structure}

The point - electron is now formally defined as composed of an electric
charge $e$ ($e\in \rm{Re}$) and two monopoles of opposite magnetic charge, 
$g$ and $-g$ ($g\in \rm{Re}$). The $g>0$ monopole is defined as a source
of magnetic field while the $g<0$ monopole as a sink for this field. By this
definition the covariant equations for the electric and magnetic electron's
components are: 
\begin{eqnarray}
\partial _{\mu }F^{\mu \nu } &=&-j_{e}^{\nu } \\
\partial _{\mu }F_{D}^{\mu \nu } &=&j_{g}^{\nu }  \nonumber
\end{eqnarray}

where $F^{\mu \nu }$ is the electromagnetic field tensor connected to the
four - vector $j_{e}^{\nu }$ formed by the electric charge density and
current, and $F_{D}^{\mu \nu }$ is the dual of $F^{\mu \nu }$ (which is
connected to the four -vector $j_{g}^{\nu }$ formed by the magnetic charge
density and current): 
\begin{equation}
F_{D}^{\mu \nu }=\frac{1}{2}\varepsilon ^{\mu \nu \sigma \delta }F_{\sigma
\delta }
\end{equation}

if $\varepsilon ^{\mu \nu \sigma \delta }$ is the four - dimensional Levi -
Civita tensor ($\varepsilon _{0123}=-\varepsilon ^{0123}=1$, completely
antisymmetric).

Each charge and monopole can be described by a corresponding Lorentz's
equation, 
\begin{eqnarray}
m_{e}\frac{d^{2}z^{\mu }}{ds^{2}} &=&e\frac{dz^{\nu }}{ds}F_{\mu \nu }(z) \\
m_{g}\frac{d^{2}x^{\mu }}{ds^{2}} &=&-g\frac{dx^{\nu }}{ds}F_{D\mu \nu }(x) 
\nonumber
\end{eqnarray}

where $m_{e,g}$ is the (electric, magnetic) charge's inertial mass, $z_{\mu
}(s)$ is the charge's world - line as function of the proper time $s$ and $%
x_{\mu }(s)$ the corresponding four - coordinates which account for the
monopole's position.

It is important to define a valid local action principle for an electron
which is described by charges and monopoles as well as for a system of two
distinct electrons of the kind. As the electron has a half - integer spin
(to be demonstrated later) and no net magnetic charge, all the problems
related to the definition of an action principle are automatically solved
[3, 4]. The fundamental question is the formal definition of the structure,
i.e., of having a charge and monopoles at the same spacetime point by the
present definition of the structure of the electron.

The question about the possibility for a local action principle is based on
the fundamental interaction between an electric charge and a magnetic
monopole [3, 4]. Considering a system composed of just one charge and
one monopole, the fields that appear in the equations of motion (3)
represent radiative reaction terms and the fields due to their mutual
interaction. In the case of the charge, the tensor field has a term due to
the monopole influence, $(-1)\{\frac{1}{4}\varepsilon ^{\mu \nu \rho \lambda
}\varepsilon _{\rho \lambda \sigma \delta }F^{\sigma \delta }\}$, which is
the dual of the dual field tensor times $(-1)$, i.e., $(-1)(-F^{\mu \nu })$.
The $(-1)$ factor is to be multiplied by the dual of $V^{\mu \nu }$ (if now $%
\partial _{\mu }V^{\mu \nu }=j_{g}^{\nu }$) when considering the action upon
the electric charge.

Omitting the self - reacting free field of the charge on itself, the
equation of motion can be derived from the action principle on: 
\begin{equation}
m_{e}\int \left( -\frac{\partial z^{\mu }}{\partial t}\frac{\partial z_{\mu }%
}{\partial t}\right) ^{1/2}dt-e\int \frac{\partial z_{\mu }}{\partial t}%
V_{D}^{\mu }(z)dt
\end{equation}

where $V_{D}^{\mu }$ is the dual of a four - vector potential related to the
magnetic density of charge and current, 
\begin{equation}
{\partial}_{\nu}{\partial}_{\nu} V^{\mu }=j_{g}^{\mu }.
\end{equation}

Once there are no longer homogeneous field equations, potentials can no
longer be introduced by expressing the field tensor $V^{\mu \nu }$ as the
curl $V^{\mu \nu }=\partial ^{\mu }V^{\nu }-\partial ^{\nu }V^{\mu }$. The
only way it can be done is when $V_{D}^{\mu }(z)$ is never to be considered
at the points where $j_{g}^{\mu }\neq 0$ i.e., it must be assured the charge
and the monopole never met at the same spacetime point. This requirement is
not physical for the case of free monopoles, and therefore it is important
to show this can be physically achieved in the case of the electron.

In order to prove the second term in the action of equation (4) is physical
in the case of the electron, it is necessary to consider first the
corresponding nonlocal action as a consequence of $j_{g}^{\mu }\neq 0$: 
\begin{equation}
V_{D}^{\mu }(x)=\int_{-\infty }^{0}V_{D}^{\alpha \beta }(\xi )\frac{\partial
\xi _{\beta }}{\partial x^{\mu }}\frac{\partial \xi _{\alpha }}{\partial
\tau }d\tau
\end{equation}

where the four- vector $\xi ^{\mu }(x,\tau )$ is characterized by 
\begin{eqnarray}
\xi ^{\mu }(x,0) &=&x^{\mu } \\
\lim_{\tau \rightarrow -\infty }\xi (x,\tau ) &=&\rm{space-like-infinity} 
\nonumber
\end{eqnarray}

The path from space-like infinity to $x$ is traversed by $\xi $ as $\tau $
varies from $-\infty $ to $0$ and then, from equation (6) one finds: 
\begin{equation}
\partial ^{\mu }V_{D}^{\nu }-\partial ^{\nu }V_{D}^{\mu }=V_{D}^{\mu \nu
}+\int_{-\infty }^{0}\frac{\partial \xi _{\beta }}{\partial x^{\nu }}\frac{%
\partial \xi _{\gamma }}{\partial x^{\mu }}\frac{\partial \xi _{\alpha }}{%
\partial \tau }(\partial ^{\gamma }V_{D}^{\alpha \beta }+\partial ^{\alpha
}V_{D}^{\beta \gamma }+\partial ^{\beta }V_{D}^{\gamma \alpha })d\tau
\end{equation}

with $(\partial ^{\gamma }V_{D}^{\alpha \beta }+\partial ^{\alpha
}V_{D}^{\beta \gamma }+\partial ^{\beta }V_{D}^{\gamma \alpha })=\varepsilon
^{\alpha \beta \gamma \sigma }\left[ j_{g}\right] _{\sigma }(\xi )$. The
action will be nonlocal as long as, in order to get $\partial ^{\mu
}V_{D}^{\nu }-\partial ^{\nu }V_{D}^{\mu }=V_{D}^{\mu \nu }$ from equation
(8), the paths of the charge and monopole are previously set to not cross,
so that a charge never meets a monopole at the same spacetime point.

Besides the magnetic charge in the point - electron case is null (as it is
composed of two monopoles of opposite sign plus an electric charge) it is
important to define the conditions a monopole and a charge can be considered
at the same spacetime point , i.e., it must be verified to be valid for the
particular case of the electron's structure.

Considering the general case of a charge and a monopole that meet at some
spacetime point, the contribution for $F^{\mu \nu }$ in equation (3) as
given by the second term on the right of equation (8) will be [3]: 
\begin{equation}
\lim_{\tau \rightarrow 0}-\left[ \frac{\partial \xi _{\alpha }}{\partial
\tau }\right] _{\tau =0}\varepsilon ^{\alpha \mu \nu \sigma }\int_{\tau
}^{0}\left[ j_{g}\right] _{\sigma }(\xi )e\frac{dz_{\nu }}{ds}d\tau .
\end{equation}

This contribution will vanish provided 
\begin{equation}
\frac{dz_{\nu }}{ds}=\frac{dx_{\nu }}{ds}
\end{equation}

(according to the definitions in equation (3) $j_{g}$ is expressed in terms
of $dx_{\nu }/ds$), as in this case $\varepsilon ^{\alpha \nu \mu \sigma
}(dz_{\sigma }/ds)(dx_{\nu }/ds)=0$. Therefore the local character of the
action principle is assured as long as, when at the same spacetime point,
the charge and the monopole have the same four - velocities. This of course,
is the case for the electron's structure, then the concept of having the
electric and magnetic charges at the same spacetime point is well defined
and leads to a valid local action principle for theparticles in the electron's structure.

In order to have a general variational principle from which both particle as
well as field equations can be derived simultaneously for a system of
distinct electrons, it is necessary to have $j_{g}^{\mu }=0$ everywhere [3].
This is true for the case of the electron's structure as it is composed of
two monopoles of opposite sign at the same point with the charge: No net magnetic charge is defined and no free monopole is considered. As
there is \textit{also }no crossing of the world - lines of two distinct spin
one - half electrons, a variational principle exists for the complete theory.

In order to conclude the electron's structure definition it is necessary to
determine its spin. In classical or quantum approach the electron is
supposed to have a nonzero radius in order the self - energy part due to the
electromagnetic fields be finite. The stability and self - energy
problems are not the concern in this work. The important issues are the
facts the electron has a defined discrete electric charge and a quantized
intrinsic magnetic flux. If some radius is to be assigned to the electron, where by the definition all the fields are set to be zero within the
defined sphere of that radius, the charge and monopole densities inside the
sphere are unknown. The important information is that these densities exist
for some electron's radius but no net magnetic flux on closed surfaces can
be found since any closed surface must entirely contains the whole structure
(the \textit{elementary} particle has now a minimum radius). In principle,
when $g\neq 0$ is assigned on some point, a variational principle cannot be
defined in a way to derive the equations of motion for particles and fields
simultaneously. In order to calculate the electron's spin one considers the
monopoles are not at the same position together with the charge, but
arbitrarily close, then a general variational principle holds in this limit.

The electron's angular momentum can be easily calculated. It is expected by
the Dirac equations that an irregular circulatory movement of the electron's
electric charge (Zitterbewegung) is the current responsable for the
intrinsic magnetic moment. The two monopoles are disposed within the
electron's structure in a way that the intrinsic magnetic dipole is the
combination of a positive and a negative monopole. This combination will be
indistinguishable from that produced by the suitable current due to
Zitterbewegung [3].

The calculation now proceeds for the monopoles and charge system: In a
Cartesian coordinate system, consider a positive monopole $g$ at $%
(0,0,z_{0}) $ with $(z_{0}>0,z_{0}\in \rm{Re})$, a negative monopole $-g$
at $(0,0,-z_{0})$ and an electric charge $e$ at the origin. It is possible
to calculate the angular momentum of the electromagnetic field, $\mathbf{L}%
_{em} $: 
\begin{equation}
\mathbf{L}_{em}=\frac{1}{4\pi c}\int \mathbf{r\times }\left( \mathbf{E}%
\times \mathbf{B}\right) dv
\end{equation}

where $\mathbf{r}$ is the three - vector distance, $\mathbf{E}$ and $\mathbf{%
B}$ the electric and magnetic fields respectivelly, and $dv$ the volume
integration element. It results: 
\begin{equation}
\mathbf{L}_{em}=\frac{\left( 2g\right) e}{c}\frac{\mathbf{z}}{\mid \mathbf{%
z\mid }}
\end{equation}

with $\mathbf{z}/|\mathbf{z\mid }$ the versor on the positive direction on $z
$ axis. It is independent of $z_{0}$, so the monopoles can be defined 
\textit{arbitrarily} close to the charge for the same resulting angular
momentum. Once the monopoles are defined inside the electron's radius, a
magnetic dipole field is generated and no isolated monopole is considered at
all, as the combination of the two monopoles must account for the
Zitterbewegung current.

The magnetic flux is expected to be quantized for a discrete electron's
charge, due to Dirac strings or due to simple quantum - arguments ( the
quantization of the angular momentum). The Dirac quantization condition [1], 
$\mu e/c=nh/2$ ($n$: integer), with $\mu $ the \textit{total} magnetic flux,
gives the expected quantized magnetic flux. Since the resulting angular
momentum of the electric charge $e$ and the monopole $g$ is equal to that of
the same electric charge $e$ and the $-g$ monopole (as this last monopole is
opposit directioned related to the first, regarding the central electric charge) the total angular momentum is as
shown by equation (12). By Dirac quantization condition one then gets $%
\left( 2g\right) e/c=nh/2$ for the system. The same result can be achieved
if the electromagnetic field is quantized by half - integer numbers [5], so
that the same relation can be derived directly from equation (12) with this
assumption.

\section{Comments and conclusions}

In the view of Dirac's work [1, 2] the charge quantization can be now
properly stated since there is in fact at least one monopole in Nature (each
electron has at least two). It was shown that this structure has no problems to be described
through a variational principle [3, 4].

It is interesting to mention the fact that if Dirac's quantization condition 
$\mu e/c=nh/2$ is valid, the electron's magnetic moment, $e\hbar /(2mc)$ is $2gz_{0}$. As the angular momentum is $\left( 2g\right)
e/c=nh/2$, one gets:
\begin{equation}
z_{0}=\frac{e^{2}}{2\pi mc^{2}}\left( \frac{1}{n}\right) 
\end{equation}

where $m$ is the electron's inertial mass and $n$ an integer number (which
must be different from zero in the case of the spin electron). We get
roughly,
\begin{equation}
z_{0}\sim \frac{10^{-14}}{n}
\end{equation}

centimeters. As $n$ may have any integer value different from zero, the size
of the electron as composed of one electric charge and two monopoles of
opposite charge is arbitrarily small ($n\rightarrow \infty $) for the same
value of the magnetic moment. Once it is possible to set $z_{0}\rightarrow 0$
the variational principle for particles and fields is well defined as for
any action on the monopole $+g$, there is a $-g$ monopole arbitrarily close.
It is then interesting that in this semi - classical description the
electron as composed of classical particles may have any small radius for
some fixed known magnetic moment (in the usual classical approaches electric currents required some finite spacial dimension to be defined as source for the magnetic moment).

It is important to mention the Dirac strings associated to the monopoles had
an special attention in the past years [6, 7] as a separeted issue. Wu and
Yang get rid of the singular string influences via a Fiber Bundle formalism
in which it is not observable, [6], being considered only as a gauge
artifact. He, Qiu and Tze [7] argumented it is impossible to have $g\neq 0$
in pure Quantum Electrodynamics however. It was showed [7] that the Dirac
quantization condition is related to arbitrary gauge couplings in pure
Quantum Electrodynamics. In view of the present work the conclusions in
reference [7] are very instructive, because, if the electron can be described
as composed of monopoles in its structure, some physical significance to the
strings must be supplied in Physics.

\section{Acknowledgements}
I would like to thank professor H. -J. He (University of Michigan) for usefull  comments. I am in debt to professors C. O. Escobar (Unicamp), R. Tschirhart (Fermilab), Y. W. Wah (University of Chicago) and R. Ray (Fermilab) for the opportunity and hospitality at Fermilab where this work was developed. I am also gratefull to Centro Universit\'ario Salesiano de S\~ao Paulo Unisal for the finantial support.

\section{References}
.
\par [1] P. A. M. Dirac, Proc. Roy. Soc. \textbf{A133} (1931) 60

[2] P. A. M. Dirac, Phys. Rev. \textbf{74} (1948) 817

[3] F. Rohrlich, Phys. Rev. \textbf{150} (1966) 1104

[4] D.Rosenbaum, Phys. Rev. \textbf{147} (1966) 891

[5] M. N. Saha, Phys. Rev. 75 (1949) 1968

[6] T. T. Wu and C. N. Yang, Phys. Rev. \textbf{D12} (1975) 3845

[7]  H. -J. He, Z. Qiu and C. -H. Tze, Zeits. Phys. \textbf{C65} (1995), 
\par \textbf{hep-ph/9402293}

\end{document}